# Molecular dynamics simulation of nanofilament breakage in neuromorphic nanoparticle networks

Wenkai Wu [1], Theodoros Pavloudis [1,2], Alexey V. Verkhovtsev [3], Andrey V. Solov'yov [3], Richard E. Palmer [1]

[1] Nanomaterials Lab, College of Engineering, Swansea University, Fabian Way, SA1 8EN, Swansea, UK; 987513@swansea.ac.uk (W.W), R.E.Palmer@Swansea.ac.uk (R.P)
[2] School of Physics, Faculty of Sciences, Aristotle University of Thessaloniki, GR-54124 Thessaloniki, Greece; tpavlo@auth.gr (T.P)
[3] MBN Research Center gGmbH, Frankfurter Innovationszentrum Biotechnologie, Altenhöferallee 3, 60438 Frankfurt am Main, Germany; verkhovtsev@mbnexplorer.com (A.V), solovyov@mbnresearch.com (A.S)

E-mail: R.E.Palmer@Swansea.ac.uk

**Abstract**

Neuromorphic computing systems may be the future of computing and cluster-based networks are a promising architecture for the realization of these systems. The creation and dissolution of synapses between the clusters are of great importance for their function. In this work, we model the thermal breakage of a gold nanofilament located between two gold nanoparticles via molecular dynamics simulations to study on the mechanisms of neuromorphic nanoparticle-based devices. We employ simulations of Au nanowires of different lengths (20-80Å), widths (4-8Å) and shapes connecting two $Au_{1415}$ nanoparticles (NPs) and monitor the evolution of the system via a detailed structural identification analysis. We found that atoms of the nanofilament gradually aggregate towards the clusters, causing the middle of wire to gradually thin and then break. Most of the system remains crystalline during this process but the center is molten. The terminal NPs increase the melting point of the NWs by fixing the middle wire and act as recrystallization areas. We report a strong dependence on the width of the NWs, but also their length and structure. These results may serve as guidelines for the realization of cluster-based neuromorphic computing systems.

## 1. Introduction

Modern computers developed from Turing-von Neumann paradigm are more and more challenged by the higher demands from the blossoming novel computing concepts like artificial intelligence, big data, and the internet of things [1–6]. Moreover, the huge power consumption [7], the fast-increasing toxic electronic waste [8], and the approaching end of Moore's law [9] suggest a serious impasse of current computer technologies [10] and call for a new computer framework. Early in 1948, Turing predicted a computing machine which works like the neural system [11]. In such a machine, artificial neurons are all connected in an arbitrary pattern with modifiers in-between. This goal was explicitly spelled out by Mead [12] and has nowadays developed into an interdisciplinary subject called neuromorphic computing, which promises to overcome some of the limitations of conventional computers [13,14].

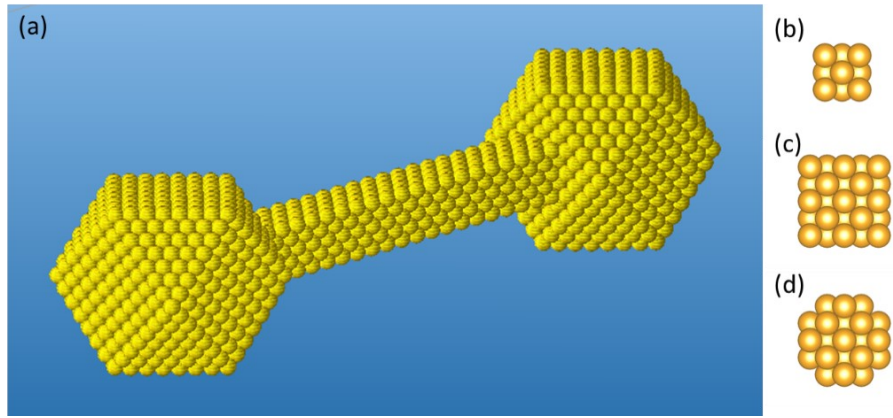

**Figure 1** (a) The structure of one of the simulated systems. An 8Å wide, 60Å long cubic NW is connected to two cuboctahedral $Au_{1415}$ clusters at each end. The cross-section of different NWs considered in this work: (b) the 4Å-wide cubic NW (c) the 8Å-wide cubic NW (d) the 8Å-wide truncated cubic NW.

One of the most promising architectures of neuromorphic computers is based on assembling interconnected, synapse-like switching devices [15–17]. The main elemental building block to act as a neuron in these devices is called a memristor [18,19], a non-linear two-terminal electrical component, which changes its resistance depending on the history of bias applied to it [20]. One of the important ways to achieve this attractive feature is the formation and dissolution of nanoscale conductive filaments [21]. Depending on the specific application, the formed conductive filaments have a lifetime (or retention time) from few microseconds [22,23] to several months or even years [24,25], which can mimic both short-term and long-term plasticity in neuron network. Research shows that this process is strongly affected by surface atomic self-diffusion [26].

Lithographic methods are commonly used in fabricating memristors [27–29], but they are complicated and costly due to the need to deterministically create robust intra- and inter-device connections [30]. A simpler and possibly cheaper alternative is the random assembly of nanoscale building blocks to form the neuromorphic network from the bottom-up [30]. In such an approach, nanoparticles (NPs) can be used as building blocks for memristors [16,31–34]. NP films produced by gas-phase cluster deposition technology have been reported to show non-ohmic electrical behaviour and reproducible resistive switching [33–36]. Such films have been deposited on a multitude of insulating materials, such as silicon nitride [34,35], silica [33], glass [30] or even paper [36]. Networks fabricated via this method show good robustness because their overall structure is fixed on the substrates [37], in contrast to some other approaches like annealing thin film to form clusters [38].

The mechanisms of formation and breakage of the conducting connections between NPs are of utmost importance and intensive experimental research and simulations have been done on filamentary processes. In general, atomic-scale connections between the gaps in the NP film are created by the electric field induced surface diffusion and evaporation [37] and also by van der Waals forces between metal atoms. After its creation, this connection introduces electromigration which eventually causes its breakage. There are numerous studies on the electromigration contribution to the breakage or resistive switching of the nanowires (NWs) [39–42]. One alternative mechanism for the formation and dissolution of the conductive paths between particles is due to the Joule heating and temperature alternation caused by the variation of the current flow [31]. When a point contact in a complex nanoscale junction is smaller than the mean free path of electrons, high temperature up to thousands of K could be induced by a small voltage drops like 100 mV [43].

The thermally induced breaking mechanism of NWs has been investigated in many reports [44–48]. Experiments show that large local heat can be generated in a NW network by the current [49] and the breaking process by the heat generation has been observed in-situ with TEM [50]. The effect of Joule heating on the breaking process of connections between particles in percolating systems is non-negligible, but there is a shortage of such studies.

In this study, we performed molecular dynamic (MD) simulations on the breakage of nanofilament to verify the Joule heating contribution to the breaking mechanism. Gold NWs with different lengths, widths and structures were located between two nanometer-sized gold clusters in order to mimic a transit nanofilament. The bottom atoms of the gold clusters are fixed to imitate the interaction with the substrate. The temperature was gradually raised up to find the breaking condition of each simulated system. Structural changes in the system were analysed. Our results show that the connection between NPs can be broken by Joule heating, validating the mechanism in cluster-based neuromorphic devices.

## 2. Methodology

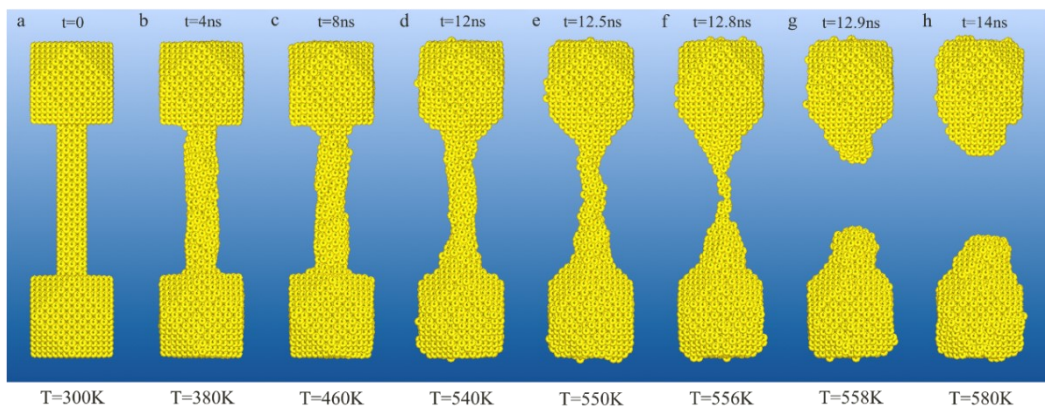

**Figure 2** The breaking process of an 8Å wide, 60Å long cubic NW between two Au$_{1415}$ clusters. The initial temperature is 300K (a). Panels (b-d) illustrate the perturbation region: some atoms of the NW are gradually pulled towards the clusters, and the NW's diameter is gradually reduced. In panels (e) and (f), the atoms of the NW move towards the clusters on two sides more quickly. The NW becomes even thinner in the middle. Fragmentation of the NW is shown in panel (f). After the breaking, the atoms of the NW coalesce into the NPs (g-h).

The MD simulations were performed with the MesoBioNano Explorer (MBN Explorer) software package [51]. The creation of the systems and analysis of the results were performed with its accompanying multi-task toolkit, MBN Studio [52].

The simulated systems consist of a gold NW as the nanofilament connecting two Au$_{1415}$ cuboctahedral clusters [53]. The atoms on the bottom facet of the clusters were fixed to simulate the bonding with the substrate. The NWs have a face-centred-cubic (fcc) structure. The joints of the NW and the clusters are at the central axis of the system on the (100) facet. Figure 1,a shows the structure of one of the simulated systems. Three kinds of NWs were simulated: a 4Å-wide cubic NW (3 layers of atoms), an 8Å-wide cubic NW (5 layers of atoms) and an 8Å-wide truncated cubic NW, in which the atoms on the four long edges of the NW have been removed. These shapes were chosen to reproduce experimental observations of Au NWs of a truncated cubic shape [54–56]. The cross-sections of the three NWs are shown in Figure 1,b-d.

We used the Gupta many-body potential [57] for our MD simulations. The parameters for Au-Au interaction were adopted from Ref. [58] with a cutoff of 7Å. The Langevin thermostat with a 1 ps damping time was used to control the temperature, and the timestep for the MD was set at 2 fs.

At first, all the models were relaxed using the conjugate gradient algorithm to obtain the lowest energy configuration at 0K. Then, the MD simulations were initialised at room temperature (300 K), and the models were heated at a rate of 20 K/ns for 30 ns. Each model was simulated seven times with random initial microstates [59] to reflect the statistical nature of the NW breakage process. The trajectory, the temperature and the potential energy changes for each simulation were recorded. To extract the crystallinity of the NWs and NPs during the heating process, an adaptive Common Neighbor Analysis (a-CNA) [60] was performed using the OVITO visualization and analysis software [61].

### 3. Results

#### 3.1. Breakage process of the nanofilament

The breaking process of an 8Å wide, 60Å long cubic NW between two Au$_{1415}$ clusters is illustrated in Figure 2. The typical evolution of the NW in the simulated systems can be summarised in the following stages: At low temperatures, the shape of the NW is changed minimally due to the thermal vibrations of the system. The NW loses its crystallinity and its outer most atoms begin to diffuse towards the clusters in a slow and gradual manner (Figure 2, b-d). As the temperature rises, the atomic movement towards the clusters becomes more pronounced, leaving the middle part of the NW thinner and more unstable (Figure 2, e, f). Eventually, the NW breaks, and all the atoms left in the middle are pulled towards the clusters quickly and are assimilated into them. Figure 2, f and g show the two trajectory frames before and after the breaking event in our simulation. After the breaking, the aggregation continues but in a much slower manner (Figure 2, h) until the whole system melts and becomes amorphous.

In this study, we consider fragmentation of a nanowire on the atomistic level by means of the MD approach. This imposes limitations on the sizes of the studied systems, which are much smaller than the systems simulated in Ref. [47,48] by means of the kinetic Monte Carlo method. The cited studies demonstrated that micrometer-sized, metal NWs can break thermally into smaller islands which then and coalesce, a phenomenon analogous to the Rayleigh instability for a liquid cylinder. In our simulations, the NWs always break at the middle, and not disperse into multiple islands. This is caused

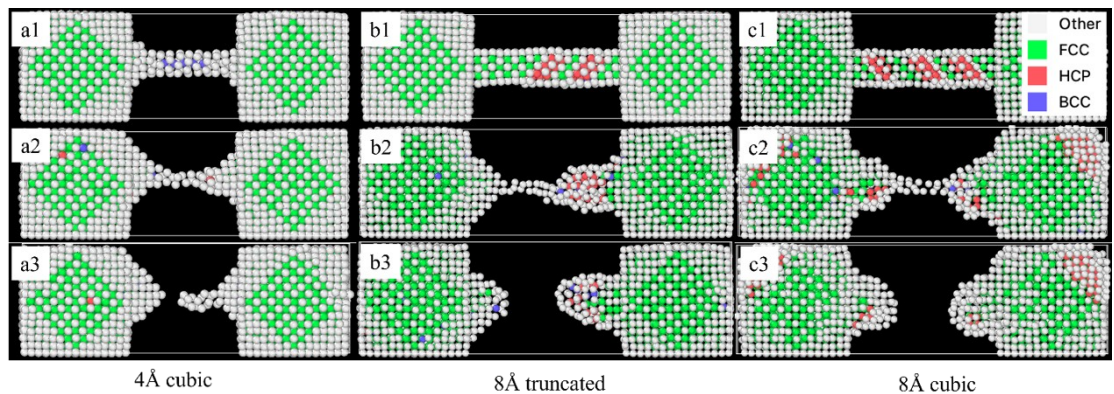

**Figure 4** (a1-c3): A comparison of the melting process between the systems with different NWs. Columns: (a) The 4Å-wide 40Å-long cubic NW. (b) The 8Å-wide 40Å-long truncated cubic NW. (c) The 8Å-wide 40Å-long cubic NW. Rows: (1) The frame during the perturbation period at t=1ns. (2) The frame before the NW breaks. (3) The frame after the NW breaks.

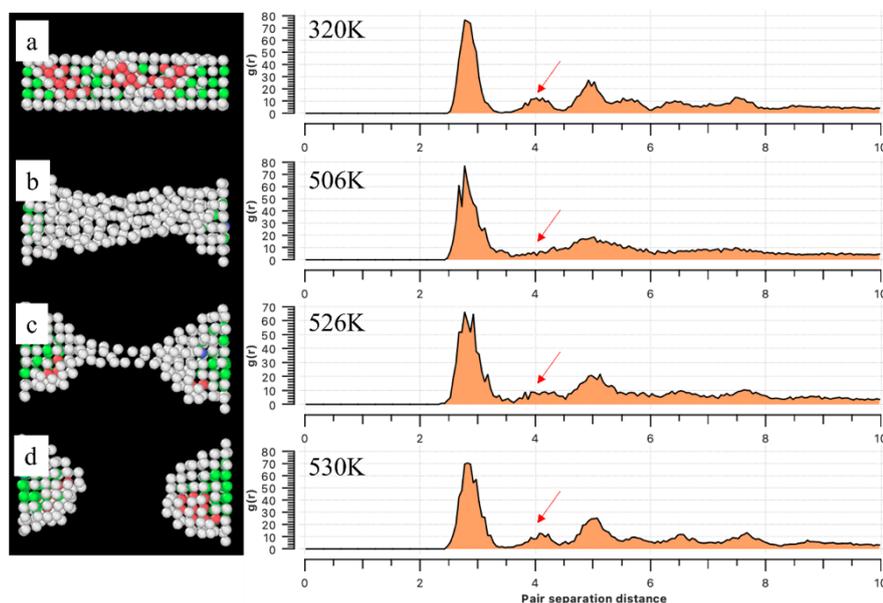

**Figure 4** (a-d): The NW part of the 8Å-wide 40Å-long cubic NW model and corresponding radial distribution functions (a) The NW at the perturbation region. (b) The $\sqrt{2}\sigma$ peak is reduced and (c) reaches to its minimum. (d) The $\sqrt{2}\sigma$ peak reappears due to recrystallisation at the two NPs.

by the strong surface diffusion current towards the clusters, which reduces the thickness of the NW in a gradual way.

## 3.2. Structural analysis

A structural analysis can reveal further details of the mechanism of the breaking of the NW. Figure 3 shows the results of this analysis for some critical frames of the breaking process of NWs with different shapes and widths. The atoms of the inner shells of the $Au_{1415}$ clusters and the connecting NWs are identified being in the fcc structure initially. The atoms of the outer shells, on the other hand, cannot be assigned by the CNA algorithm to any of the crystalline structures due to their reduced coordination number and are therefore classified by the algorithm as "amorphous", despite them being visibly arranged in a crystalline formation.

In the thinner 4Å-wide cubic NW (see Fig. 3(a1-a3)), the surface atoms constitute a large proportion of the whole NW, therefore only a single row of atoms in the center appears crystalline from the beginning. The algorithm can hardly show the structure change of the NW in this case. However, as more atoms aggregate towards the clusters, atoms marked as fcc or bcc will appear on both ends of the NW.

For the 8Å-wide cubic NW (Fig. 3(c1-c3)) and 8Å-wide truncated cubic NW (Fig. 3(b1-b3)), the structural changes during the breaking process are similar. During the perturbation stage, phase transitions from fcc to hcp structure appear diagonally at 2 and 3 places, respectively, dividing the NWs in equally sized parts. As the temperature rises and more atoms aggregate towards the NPs, the structural transitions between fcc and hcp become more frequent and irregular. The middle part of the NW becomes too thin for the algorithm to recognise the structure, but inner atoms at the ends of the NW remain crystalline. These phase transitions are caused by the shearing strength generated by the perturbation. Marszalek et. al. [62] confirmed experimentally that the sliding of crystal planes within the gold NWs will change the local structure from fcc to hcp. This phenomenon is known for plastic deformation in fcc metals.

**Table 1** The average breaking temperature of the different systems

| NW width and structure / NW length | 20Å | 40Å | 60Å | 80Å |
|---|---|---|---|---|
| 4Å cubic | 448±29K | 434±32K | 444±32K | 469±25K |
| 8Å truncated cubic | 556±52K | 484±15K | 491±19K | 494±18K |
| 8Å cubic | unbreakable | 543±13K | 516±30K | 493±64K |

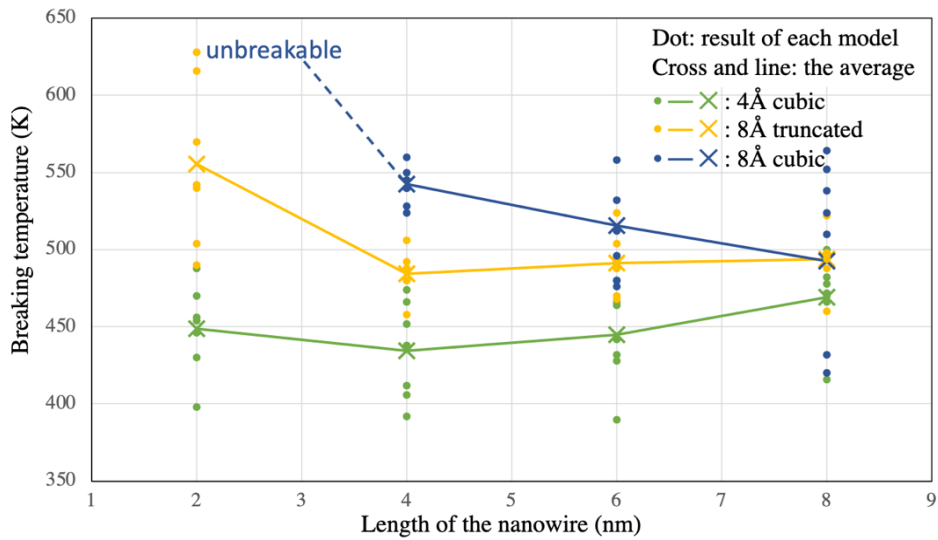

**Figure 5** The breaking temperatures of all the individual systems (dots) and the average values of the breaking temperature for the specific length and width of the NW (crosses and lines).

In the studied systems, parts of the NW, specifically their two ends, remain crystalline during the whole process, a finding that differs from previous studies of the melting of fcc NWs [44,45]. This is owed to the presence of the clusters at the ends, which act as a recrystallisation area. The amorphisation of the NW and re-crystallisation at the two ends also visible by the radial distribution function. Figure 4 shows snapshots of 8Å-wide 40Å-long cubic NW between two clusters and the corresponding radial distribution function. The nearest neighbour distance ($\sigma \sim 2.8$Å) can be read from the first peak. Figure 4,a shows a crystalline state. During the perturbation region of the NW, the $\sqrt{2}\sigma$ peak of the function begins to decrease (Figure 4,b). The $\sqrt{2}\sigma$ peak reaches its minimum (Figure 4,c) and the form of the functions tends resembles a liquid state. In Figure 4,d, the NW is broken and the $\sqrt{2}\sigma$ peak gradually appears again. In general, the disappearance of the $\sqrt{2}\sigma$ peaks of happens ~10÷30K before the breaking temperatures.

### 3.1 The breaking temperature

The average breaking temperatures of the simulated systems with the different lengths and widths of the NWs are shown in Table 1 and illustrated in Figure 5. The breaking temperatures in most cases vary from 400K to 650K and show a dependence on the width, length and structure of the NW.

Wider NWs will generally break at higher temperatures and this becomes more pronounced when the NWs are shorter. The breaking points of 4Å- and 8Å-wide 80Å-long cubic NWs differ by ~25K. This difference however is increased to over 70K for a NW length of 60Å and to 100K for 40Å. The breaking point's dependence on the length varies according to the NW width and structure.

There is an exponential decrease of the breaking temperature with the increasing length for the 8Å NWs, with this behaviour being more pronounced for the cubic NW. The breaking point of the truncated NW remains the same for lengths ⩾40Å (±10K). The 4Å NWs always break at more or less the same temperature (±25K) regardless of their length. The slight increase in the breaking temperature of the 4Å-wide cubic NW is within the expected fluctuations in an MD simulation. Finally, cubic NWs are more robust than the truncated ones of the same width for lengths ⩽60Å and the difference is increasing as the length is being reduced. However, they appear to break at the same temperature at a length of 80Å.

The 8Å-wide 20Å-long cubic NW is an interesting case, as it retained its integrity during the whole simulation, until the whole the system ended in one completely melted amorphous structure at about 850K.

Experimental studies [48] show a relation between the breakup temperature of NWs and the corresponding bulk melting temperature of the material given by:

$$\frac{T_{mn}}{T_{mb}} = C_1 - \frac{C_2}{d_{NW}} \qquad (1)$$

where $T_{mn}$ is the observed breaking temperature of the NW and $T_{mb}$ is the melting temperature of the corresponding bulk material. $C_1$ and $C_2$ are constants related to the material and $d_{NW}$ is the initial diameter of the NW. For the pure gold NWs with a 4Å and 8Å diameter respectively, the predicted breaking temperatures are 120K and 290K respectively. This is a rather large difference with our simulation result, that is owed to the fact that the clusters have a large role in stabilizing the NW in the middle.

The Rayleigh instability model, which describes the breaking up of liquid cylinders, has been used to describe the fragmentation process of NWs [46]. The applicability of this model is questionable in a fragmenting NW [47], however, the behaviour observed in our MD simulations agree qualitatively with the breakage behaviour of a liquid cylinder described by means of the Rayleigh model.

We can consider a short period around the breaking event as the stable state applicable to the Rayleigh model, e.g., according to our description of the breaking process in Figure 2 f-h, the state 0.2 ns after the breaking of NW. We measure the distance $\lambda$ between the two parts of the broken NW, the diameter of the rest of the NW after the breaking event $d_{sp}$, and its initial radius $R_0$, calculated by the average distance between the center axis and the outer atoms and calculate the $\lambda/R_0$ and $d_{sp}/R_0$ ratios for the more cylindrical 8Å-wide truncated cubic NW are 8.2 and 2.5, respectively. Nichols and Mullins predicted $\lambda/R_0$ and $d_{sp}/R_0$ values equal to 8.89 and, assuming constant volume, 3.76 as the criteria for surface diffusion dominated breakup of a cylindrical wire as result of harmonic surface perturbations [63].

Thus, our results are in good agreement with the Rayleigh model suggesting that surface diffusion dominates the process. Deviations from the theoretical predictions are to be expected since our simulated system is still not of the ideal shape and isotropy which are assumed in the Rayleigh model with atoms of the NW diffusing towards the clusters on the sides.

## 5. Conclusions

In an effort to gain insight on the breaking process of nanoflilaments connecting clusters, we built a series of model systems and performed MD simulations on the breaking process by heating the system to imitate the Joule heating mechanism in a cluster-based percolation system and analysed the evolution of these systems.

Atoms on the NW will gradually aggregate towards the clusters on the sides, and the NW will eventually break at a certain temperature before the melting of the whole system. The a-CNA structural analysis result showed that large part of the NW would keep crystalline, but the middle part of the NW will be molten during the breaking. The breaking temperature's dependency on the geometric characteristics of the NWs was extracted. We find a strong dependence on the width and structure of the NW, while the length is found to be an important parameter in some, but not all cases.

The NWs breaking temperatures determined from the MD simulations presented in this study are much higher than the expected breaking temperatures for isolated NWs of the same diameter. We attribute that to the clusters on the sides which stabilize the NW in the middle. The Rayleigh model can be applied to analyze the molten part of the NW, which indicates that the breaking mechanics of the NW in the simulated systems is a consequence of diffusion processes.

We showed that nanofilaments between clusters can be broken by Joule heating. Since structures similar to our model systems are applied in cluster-based neuromorphic system, our work indicates that the Joule heating mechanism is likely to be an important factor in the breaking process of the percolation connections of these systems. We remark that a theoretical model needs to be established to account for filament formation, as well as breaking. We hope that this work will contribute towards that goal.

## Acknowledgements

The authors are grateful for partial financial support from the European Union's Horizon 2020 research and innovation programme -- the RADON project (GA 872494) within the H2020-MSCA-RISE-2019 call. This work was also supported in part by Deutsche Forschungsgemeinschaft (Project no. 415716638).

We acknowledge the computational support by the Supercomputing Wales project, which is part-funded by the European Regional Development Fund (ERDF) via the Welsh Government, and Mr. T. D. Pritchard on using the Sunbird supercomputer.